# Highly tunable lateral homojunction formed in 2D layered $CuInP_2S_6$ *via* in-plane ionic migration

*Huanfeng Zhu[1,2]\*, Jialin Li[1], Qiang Chen[1], Wei Tang[1], Xinyi Fan[1], Fan Li[1], and Linjun Li[1,2]\**

1. State Key Laboratory of Modern Optical Instrumentation, College of Optical Science and Engineering, Zhejiang University, Hangzhou 310027, China.

2. Intelligent Optics & Photonics Research Center, Jiaxing Research Institute Zhejiang University, Jiaxing 314000, China

\*Corresponding Author. E-mail: hfzhu@zju.edu.cn (H.Z.); lilinjun@zju.edu.cn (L.L.)

As basic building blocks for next-generation information technologies devices, high-quality *p-n* junctions based on van der Waals (vdW) materials have attracted widespread interest. Compared to traditional two dimensional (2D) heterojunction diodes, the emerging homojunctions are more attractive owing to their intrinsic advantages, such as continuous band alignments and smaller carrier trapping. Here, utilizing the long-range migration of $Cu^+$ ions under in-plane electric field, a novel lateral *p-n* homojunction was constructed in the 2D layered copper indium thiophosphate (CIPS). The symmetric Au/CIPS/Au devices demonstrate an electric-field-driven resistance switching (RS) accompanying by a rectification behavior without any gate control. Moreover, such rectification behavior can be continuously modulated by poling voltage. We deduce that the reversable rectifying RS behavior is governed by the effective lateral build-in potential and the change of the interfacial barrier during the poling process. Furthermore, the CIPS *p-n* homojuction is evidenced by the photovoltaic effect, with the spectral response extending up to visible region due to the better photogenerated carrier separation efficiency. Our study provides a facile route to fabricate homojuctions through electric-field-driven ionic migration and paves the way towards the use of this method in other vdW materials.

## 1. Introduction

The flourish of various van der Waals (vdW) materials and the need for the device miniaturization stimulate rapid growth of interest in the fabrication of vdW-based *p-n* junctions, which have great potential in the modern functional diversity electronic and optoelectronic devices.[1, 2] In this direction, significant efforts have been devoted to develop the *p–n* heterojunctions.[3, 4] However, lattice mismatch-induced impurities[5, 6] and discontinuous band alignments[7, 8] are inevitably introduced at the interface which deteriorate the quality and performance of device. Compared with heterojunction, the homojunction, which is achieved through the modulation of charge type and density in different regions of same material, can effectively solve these problems.[9-14] Till to date, the conventional routes to construct lateral *p-n* homojunctions in 2D vdW material include laser irradiation,[15, 16] plasma doping[14] and layer-engineering.[17, 18] However, these approaches usually involve a complicated device fabrication process and tend to suffer from chemical contamitation. Therefore a controllable contamination free method is highly desired.

The emerging 2D vdW material CIPS has attracted intensive attention owing to its multifunctionality behavior, including ferroelectricity,[19, 20] ionic conductivity,[21, 22] pyroelectricity,[23] electrocaloric effect[24] and giant negative piezoelectricity.[21] The $Cu^+$ ion position and the migration behavior in the sublattice play a crucial role for the control of these characteristics. For instense, it has been reported that the ferroelectricity is attributed to the displacement of Cu ions along the out of plane (OOP) direction[19, 25] and the intralayer migrations of $Cu^+$ ions can induce ferroelectric domain inversion (see Figure 1a).[26] Particularly, previous calculations show that the hopping/motion barrier for $Cu^+$ ions migration in the in-plane (IP) direction ($E_a \approx 0.23$ eV) is even lower than that in the OOP direction ($E_a \approx 0.85$ eV).[22, 27] Since the $Cu^+$ ions can hop along the IP direction, it is reasonable to infer that the CIPS may show great potential in forming ionic migration induced lateral homojunction (see Figure 1c). Additionally, gaining further insights into the ion migration behavior and electron transfer at the interface of these materials could shed light on the miniaturization of functional devices.

In this work, inspired by the long-range migration of $Cu^+$ ion, a novel lateral homojunction based on layered CIPS is demonstrated. The *p-n* homojunction exhibits an obvious diode-like RS behavior with rectification characteristics can be continuously modulated. The $Cu^+$ ion migration under the in-plane electric field affects the surface potential of CIPS and the height of the interface barrier, resulting a reversible rectifying RS behavior in the proposed *p-n* homojunction. This scenario could be further confirmed by the Raman and KPFM results. Moreover, the built-in potential can effectively separate the photogenerated carriers, and a highly enhanced photovoltaic effect is observed. Our findings demonstrate a facile way to realize an in-plane vdW homojunction through field-driven ionic processes without the need of chemical doping or complicated transfer process, which exhibits great significance in the applications of novel electronic and self-powered optoelectronic devices.

## 2. Results

For the electrical measurements, the CIPS nanoflakes were mechanically exfoliated from bulk crystal on the transparent PDMS and subsequently transferred to prepared Au electrode produced in the Au/CIPS/Au structure (see more details in "Methods" and Supplementary Figure S1). A typical current-voltage (*I-V*) curve of the device is shown in Figure 1d. As a symmetrical structure (Au/CIPS/Au) is utilized, the *I–V* curve is symmetrical with polarity. At the first stage, with a negative bias voltage swept from 0 to -8 V, the device is triggered from high resistive state (HRS) to low resistive state (LRS) at a threshold voltage ($V_{th}$) of about -6.0 V. However, when the negative voltage sweeps back, the device returns to the HRS. The hysteresis loop indicates that the device exhibits bidirectional threshold resistive switching (RS) effect,[28] demonstrates that it can be developed for selector,[29, 30] logic application[31] and neuromorphic computing.[32] Similar results were also observed in few-layer SnS, which *I–V* characteristics were dominated by the in-plane ferroelectricity of the SnS.[33] However, in this work, the mechanism is quite different, which will be discussed below. The *I-V* curves in the low field (±5 V) with different sweep directions are plotted in the inset of Figure 1d. During the backward sweep, the *I-V* curve exhibits a forward diode behavior, whereas the forward sweep displays a backward diode behavior, demonstrating the reversible diode-like behaviors. Interestingly, as depicted in Figure 1e, an asymmetric I-V curve was observed when the voltage was swept from 0→+8 V→-8 V→0, showing the typical behavior of dominant ionic conduction.

To investigate the RS characteristics further, the *I-V* response with different sweep voltage $V_{max}$ ranging from ±10 to ±50 V has been applied to the planar Au/CIPS/Au device and the results are shown in Figure 2a. At the beginning, the *I-V* curve is almost symmetric and the switchable bidirectional diode behavior accompany with RS are also observed when the sweep voltage is smaller than ±20 V. The switching ratio $I_{LRS}/I_{HRS}$ reaches over 150 at around 4 V as $V_{max}$=20 V (see Figure 2b and c). Interestingly, as the $V_{max}$ increases further (more than ±30 V), a clear asymmetric *I-V* is observed indicating a transition from threshold RS to self-rectification in the device. Taking the sweep range of 50.0 V as an example, the device switched from HRS to the LRS at a positive threshold voltage ($V_{th}$), and produces a huge positive RS hysteresis as the voltage is swept backward from 50.0 V to 0 V. Interestingly, it is noticeable that the threshold voltage is increase with the increase of $V_{max}$. However, as the appling voltage is backward swept from 0 → -50 V → 0, the resistance is continuously maintained at HRS, which contributes to a rectifying ratio of exceeding 100 at +7 V/-7 V as shown in Figure 2d.

Next, we turn our attention to clarify the diode-like RS mechanism in the planar Au/CIPS/Au device. As schematic shown in Figure 2e, as the positive voltage swept from 0 → 20 V → 0, $Cu^+$ ions could migrate leftward and assemble near the cathode surface, leading to $Cu^+$ ions abundance near cathode and Cu deficiency near anode. It is reported that the assembly of the $Cu^+$ ions near the cathode act as donor, whereas the $Cu^+$ deficiency near the anode act as acceptor.[22, 34] Therefore, an p–n homojunction is formed in the CIPS and the device switches from initial HRS to LRS (i → ii). Applying opposite electric field could drive the $Cu^+$ ions move back to the original even distribution and the CIPS transforms back to the HRS state (ii → iii). An inverted p–n homojunction is established by further increasing the opposite electric field and the device transforming into the LRS (iii → iv), exhibits the bidirectional threshold RS behavior. However, when applying a larger $V_{max}$ (more than 30 V), more mobile $Cu^+$ ions will be driven to the cathode and may be reduced to Cu atoms. Consequently, the asymmetric interfacial barriers are formed, and result in the rectifying behavior.

Based on the above experimental facts and discussions, we attribute the electric-field-driven ionic migration to the self-rectify RS behavior of the planar Au/CIPS/Au device. To verify this scenario, Raman spectroscopy was performed at room temperature. Previous research shows that when Cu deficiencies, CIPS spontaneously segregates into $CuInP_2S_6$ and $In_{4/3}P_2S_6$ distinct chemical domains within the same crystal.[22, 35] It is reasonable to infer that this phase separation may be achieved due to the $Cu^+$ ions hopping as a high voltage is applied to the device. As shown in Figure 3a and b, in the initial state, several peaks are observed between 50~470 $cm^{-1}$. The peaks found in the 50~150 $cm^{-1}$ range correspond to the vibrational modes of the anions. The high intensity peaks around 263 and 374 $cm^{-1}$ can be attributed to the δ(S-P-S) and ν(P-P) modes, and the weak peaks at 162 and 449 $cm^{-1}$ correspond to the δ(S-P-P) and ν(P-S) modes, respectively.[36, 37] In particular, the peak at about 320 $cm^{-1}$ can be assigned to the movements of the $Cu^+$ ions within the $S_6$ octahedral voids. After the six consecutive positive/negative poling voltage cycles, the Raman spectra was performed immediately to get insight on the $Cu^+$ ions migrations. One can notice the intensity of the 320 $cm^{-1}$ peak decreases after applying negative poling voltage, attributed to the redistribution of $Cu^+$ ions. Subsequently, the intensity of the 264 and 374 $cm^{-1}$ peaks decreases, indicating the lattice distortions. Interestingly, the peak at 300 $cm^{-1}$, which attributed to anion deformation in IPS,[37] appear and increase with increasing negative poling voltage. On the contrary, after applying positive poling voltage, the 264 and 376 cm$^{-1}$ peak intensity increases while the 300 cm$^{-1}$ peak decreases, indicating Cu$^+$ ions can move back to initial site from the cathode. These phenomena are more obvious when performed the Raman spectra with applying a bias voltage, which will dramatically enhance the Cu$^+$ ions migration and result in a clear change in the Raman signals. Additionally, we have conducted spatially resolved Raman spectroscopy to trace the Cu ions migration along the in-plane direction in a new device (see Figure S2). As shown in Figure S2a, after applying a poling voltage, there were clearly some dispersed nanoparticles attached to the Au electrode. Moreover, we can observe an enhancement of Raman signal near one of electrode, which is closely related to the Cu ions migration under the in-plane electric field.

The Kelvin probe force microscopy (KPFM) allows for the measurement of the contact potential difference (CPD), work function of materials and local charge accumulation with high resolution of several nanometers. Next, we employed the KPFM to further investigate the electrical field-driven Cu$^+$ migration effect in the CIPS. The optical and AFM image of the device are shown in Supplementary Figure S3. The CPD between the AFM tip and the local surface of CIPS is defined as

$$\text{CPD}_{CIPS} = \frac{W_{tip} - W_{CIPS}}{-e}$$

where $W_{tip}$ and $W_{CIPS}$ are the work functions of the AFM conductive tip and CIPS, respectively, and $e$ is the electron charge. Therefore, the work function (surface potential) difference between two segments (i.e., near cathode and near anode) of the CIPS is given by

$\Delta$CPD= CPD$_{\text{cathode-CIPS}}$ -CPD$_{\text{anode-CIPS}}$

Figure 3c presents the CPD image of device in the initial state. In this image, one can recognize that there is no obvious surface potential difference between the cathode and anode regions. In contrast, after poling processes, the CPD image shows an obvious surface potential difference between the cathode and anode regions, as shown in figure 3e. The surface potential in the cathode regions was larger than that in the anode regions by 150 meV (see figure 3f), demonstrated that the Cu$^+$ ion migrated to the cathode, resulting in an n-doping effect near the cathode. The value of surface potential difference is smaller than the open circuit voltages (V$_{OC}$ ~ 300 meV), which will be described later, possibly due to surface states and/or some particles adsorbed on the surface of the CIPS flakes. Thus, these results indeed provides unambiguous proof that Cu$^+$ can migrate along the in-plane electric field and resulting the formation of effective lateral build-in potential in CIPS. Similar results were confirmed by repeated experiments (Figure S4-S6).

To investigate the charge transport mechanisms and interface barrier of the lateral Au/CIPS/Au homojunction, we conducted *I-V* measurement with the consecutively forward/backward voltage sweeping (0 V→±7 V) and the corresponding results are shown in Figure 4a. The current gradually increases with increasing the positive sweeping cycle and eventually reaches its saturation value, induces the device switch from HRS to LRS. The *I-V* curves with opposite voltage sweeping reproduced a similar result, manifesting rectifying behavior and the direction can be reversed by changing the sweep direction, consistent with the above results. Figure 4b shows the replotted *I-V* curve on a double logarithmic scale for the positive voltage sweeps. Overall, there are three distinctive regions that correspond to the direct tunneling (DT), Fowler-Nordheim (FN) tunneling and thermionic emission (TE). More detail about these models can be seen in the "Methods". At the initial HRS (the 1$^{st}$ forward bias sweep), due to the large barrier height at the interface of Au/CIPS, the ln($J/V^2$) depends linearly on ln($1/V$) exhibit a direct tunneling mechanism in the whole voltage range (Figure 4d and Supplementary Figure S7(a)). As mentioned above, the consecutive forward voltage can attract the Cu$^+$ ions to the cathode leading to n-type doping and band bending at the Au/CIPS interface, as schematic shown in Figure 4e. Thus, at high bias region, a linear region with a negative slope clearly appears, confirming the tunneling current follows the F-N tunneling model. By fitting the curves with Equation (4), we can extract the Schottky barrier parameter ($d\varphi_B^{3/2}$). As illustrated in Figure 4c, after 5$^{th}$ forward bias sweep, the Schottky barrier parameter is significantly decreased from 2.9 to 1.1 eV$^{3/2}$ nm due to the continuous migration of Cu$^+$ ions. With further reducing the barrier height, the carrier transport is dominated by the thermionic emission as expected. A similar trend is also observed with negative voltage sweeping and is shown in Supplementary Figure S8. Hence, the barrier height decreases as the accumulation of Cu$^+$ ions at the Au/CIPS interface, which is expected to facilitate carrier transport across the interface.

Next, we turn to demonstrate the switchable photovoltaic effect in lateral Au/CIPS/Au homojunction device. The experimental sketch is shown in Figure 5a. Due to the effective lateral built-in field in the CIPS and reduction of barrier heigh at the Au/CIPS interface, which is beneficial to promoting the separation efficiency of photogenerated charges, the zero bias transient photocurrent of the poled device is greatly higher than the initial state (see Figure 5b). Significant, the optical spectral extended at least up to 635 nm (see Figure 5c), beyond the bandgap limit of CIPS ($E_g$~2.6 eV). Moreover, the photocurrent direction is reversed by reversing the poling voltage. The *J-V* characteristics of the Au/CIPS/Au homojunction under an incident light of 375 nm at 14 W/cm$^2$ are shown in Figure 5d. The non-poled device displays no photovoltaic effect with near zero open-circuit photovoltage ($V_{oc}$) and short-circuit photocurrent ($J_{sc}$). In contrast, the poled device demonstrates typical photovoltaic effect. When poling the device with +15 V, $V_{oc}$ and $J_{sc}$ are obtained as -0.3 V and 20 mA/cm$^2$, respectively. With -15 V poling voltage, the $V_{oc}$ and $J_{sc}$ are reversed. Therefore, the above results corroborate the scenario of formation of reversible *p-n* homojunction in the CIPS layer by ion migration, while its photovoltaic response can be controlled by an external electric field.

The recent progress has also shown a photovoltaic effect in CIPS associated to the out of plane ferroelectricity at room-temperature.[38] To exclude the out of plane ferroelectricity as a possible reason for the observed photovoltaic response, we investigate the photovoltaic response as a function of temperature. The zero bias photocurrent was observed until the device has been heated up to 410 °C, much higher than the critical temperature ($T_C$ ~ 315K) of CIPS (see Supplementary Figure S9). In addition, the zero-bias photocurrent is independent of the light polarization, which is inconsistent with the photovoltaic effect scenario (see Supplementary Figure S10). However, it is reported that there also exists intrinsic IP polarization in CIPS.[20] Thus, the possible origin due to the effect of IP ferroelectricity cannot be completely excluded, further experiments to clarify this issue are still highly expected.

Finally, we turn to address the performance optimization of the device. Due to the long channel length of the current devices, it requires high drive voltage to obtain rectification, which leads to energy inefficiency and slow response time. Additionally, the formation of the Schottky barrier at the interface will also affect the ionic migration behavior. Thus, the drive voltage and response time can be improved by optimizing the CIPS channel length and choosing the appropriate electrodes and active area.

## 3. Conclusions

In conclusion, we successfully fabricated a highly tunable lateral homojunction in 2D layered CuInP$_2$S$_6$ through utilizing in-plane ionic migration. By combining the Raman with KPFM results, the formation of lateral homojunction is unambiguously confirmed, which is attributed to the Cu$^+$ ion migration. The proposed homojunction device shows a typical diode-like resistive switching behavior. Besides, the rectification behavior can be highly tuned by the external electric field through adjusting the ion dynamic within the CIPS. The surface potential difference, quantified by employing KPFM measurements, is estimated to be ~150 meV. Sweep cycle-dependent I-V measurement is presented to evaluate the carrier transport mechanisms at the interface between the CIPS channel and electrode. Specially, the photovoltaic properties of the device were found to be highly improved due to the formation of effective lateral build-in potential in the CIPS and reduction of barrier heigh at the interface. Our demonstration of homojunction through ionic migration may pave the way for novel electronic and self-powered optoelectronic applications.

## 4. Methods

*Device fabrication.* The electrode patterns were defined by standard photolithography process, and Cr/Au (8 nm/20 nm) electrodes were fabricated by electrical beam evaporation on Si substrates with a 285 nm $SiO_2$ oxidation layer. The thin CIPS flakes were obtained by mechanical exfoliation from bulk crystals transferred onto the prepared electrode through a dry transfer approach in a nitrogen-filled glove box. Finally, the whole devices were encapsulated by *h*-BN flakes.

*Device characterization.* The thickness of the flakes was identified by their optical contrast and then confirmed by atomic force microscopy. Raman spectral was carried out using a microscope confocal Raman spectrometer (WITec alpha 300R Raman system) with an excitation source of 532 nm laser at room temperature in air. The laser power was kept below 6 mW. The KPFM measurements were conducted on a Veeco Multimode Nanoscope III SPM. All electrical measurements were performed using a Keithley 2450 sourcemeter. The scan rate of the current-voltage measurements was 0.15 V/s and the voltage sweep was at 100 mV per step.

*I-V simulation.* In the thermionic emission regime, the I-V relationship is described by:[39]

$$J(V) = A^* T^2 \, exp\left(\frac{-(\varphi_B - \sqrt{q^3 V/4\pi\varepsilon_0 \varepsilon_r d})}{k_B T}\right) \quad (1)$$

where $A^*$ is effective Richardson constant, $T$ is the temperature, $\varepsilon_r$ is the permittivity of the CIPS, $\varepsilon_0$ is the permittivity of the vacuum, $k_B$ is the Boltzmann constant, $q$ is the electron charge, $d$ is the width of interface barrier, and $\varphi_B$ is the Schottky barrier. The barrier height can be calculated by the equation: $\varphi_B = -k_B T ln(J_0/AT^2)$, where $J_0$ is zero field current density, A is the emission constant.

At low bias voltage, according to the direct tunneling theory, the, current density can be expressed as:

$$J \propto V \, exp\left(-\frac{4\pi d\sqrt{2m^*\varphi_B}}{h}\right) \quad (2)$$

Where *h* and *m*\* are the Planck's constant and the effective mass of the charge, respectively. The dependence of $ln(J/V^2)$ on $1/V$ should be linear on a logarithmic scale in the direct tunneling regime.

When the applied bias is large enough, then Fowler-Nordheim tunneling dominates the current:

$$J \propto V^2 \exp\left(-\frac{4d\sqrt{2m^*\varphi_B^3}}{3\hbar qV}\right) \quad (3)$$

A plot of ln($J/V^2$)-1/$V$ will present linear relationship with a negative slope.

The Schottky barrier parameter ($\varphi_B^{3/2}d$) can be determined from the negative slope $k$ by Eqation 4:

$$k = -\frac{4\sqrt{2m^*}\varphi_B^{3/2}d}{3\hbar q} \quad (4)$$

**Acknowledgements**

This work was supported by the funding from National Key R&D Program of China (2019YFA0308602), National Science Foundation of China (general program 12174336& major program 91950205) and the Zhejiang Provincial Natural Science Foundation of China (LR20A040002).

**Conflict of Interest**

The authors declare no conflict of interest.

**Keywords**

Copper indium thiophosphate, ion migration, tunneling, photovoltaic

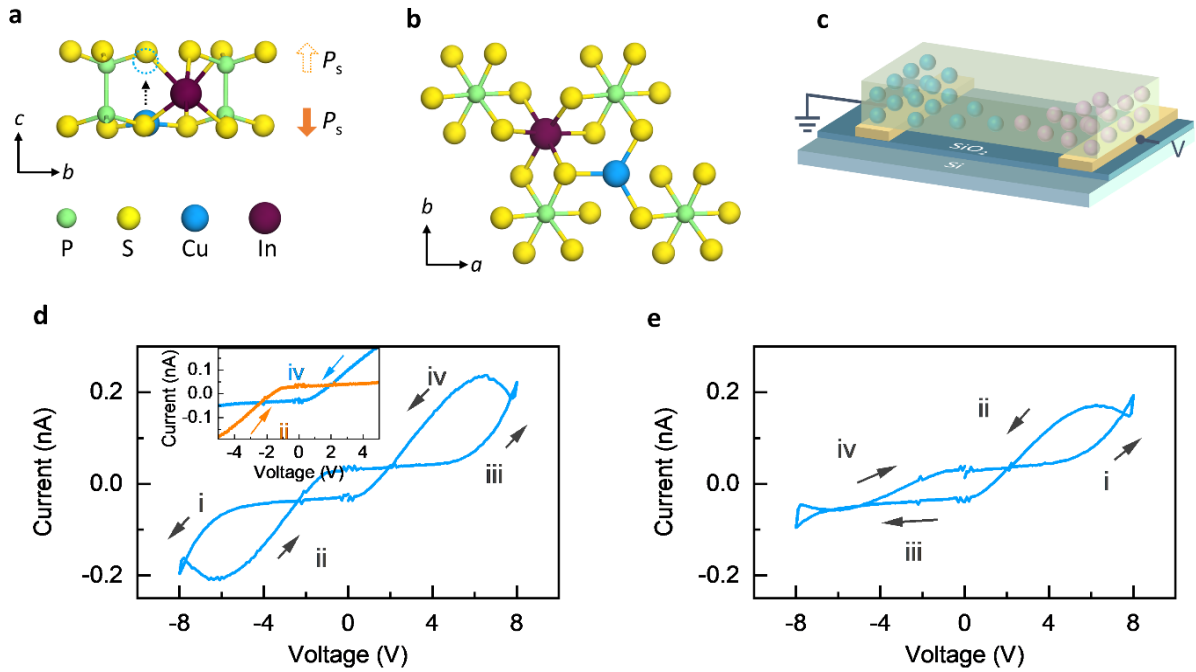

**Figure 1.** Device architecture and electrical properties. The side view (a) and top view (b) for the crystal structure of CIPS. The polarization direction is indicated in by the orange arrow, which can be inversed *via* Cu ions migration along the out of plane direction. (c) Schematic illustration of strategy for the preparation of 2D homojunctions induced by ions migration. Full *I–V* characteristic swept in the order (d) 0 V→-8 V→0 V→+8 V→0 V and (e) 0 V→+8 V→0 V→-8 V→0 V as shown by the arrows with the four sweeps labelled as i, ii, iii and iv. Insert of (d) is the replotted *I-V* segments in the range of ±5 V.

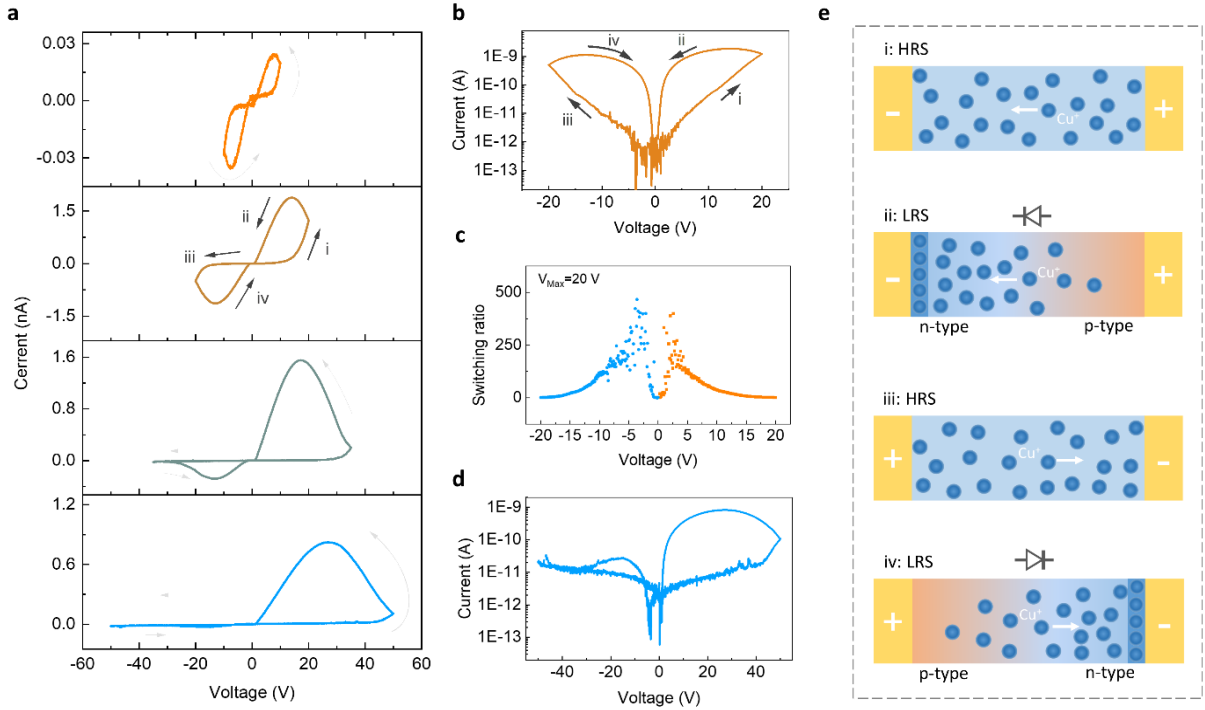

**Figure 2.** Self-rectify RS behavior and switching mechanism. (a) *I–V* curves measured with varying sweep range, $V_{max}$ is from 10.0 to 50.0 V. (b) The semilog *I-V* curve with maximum sweep voltage of ±20 V and (d) ±50 V. The arrows indicate the voltage sweep direction. (c) Switching ratio as a function of reading voltage with $V_{max}$ =20 V. (e) Schematics of the dynamic ion migration during RS switching. The i, ii, iii, and iv are four states in the *I-V* curve of (b). The white arrows indicate the direction of the $Cu^+$ ion migration driven by the applied electric field.

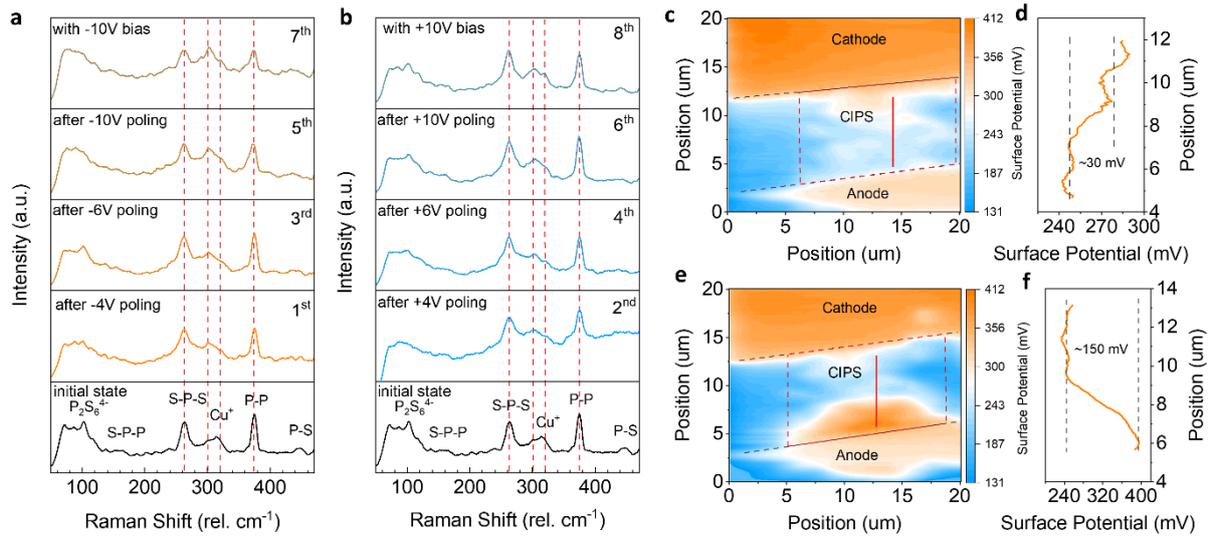

**Figure 3.** Poling voltage-dependent Raman and KPFM spectra. Raman spectroscopy obtained after six times poling voltage sweeping from 0 V to a fixed (a) positive (b) negative voltage (4, 6 and 10 V) or with ±10 V bias. The order of measurement is labeled as "$n^{th}$". Surface potential mapping of (c) initial and (e) after poling. The red dotted lines show the approximate positions of the CIPS. (d), (f) Cross-section profile along the red line in (c) and (e), respectively.

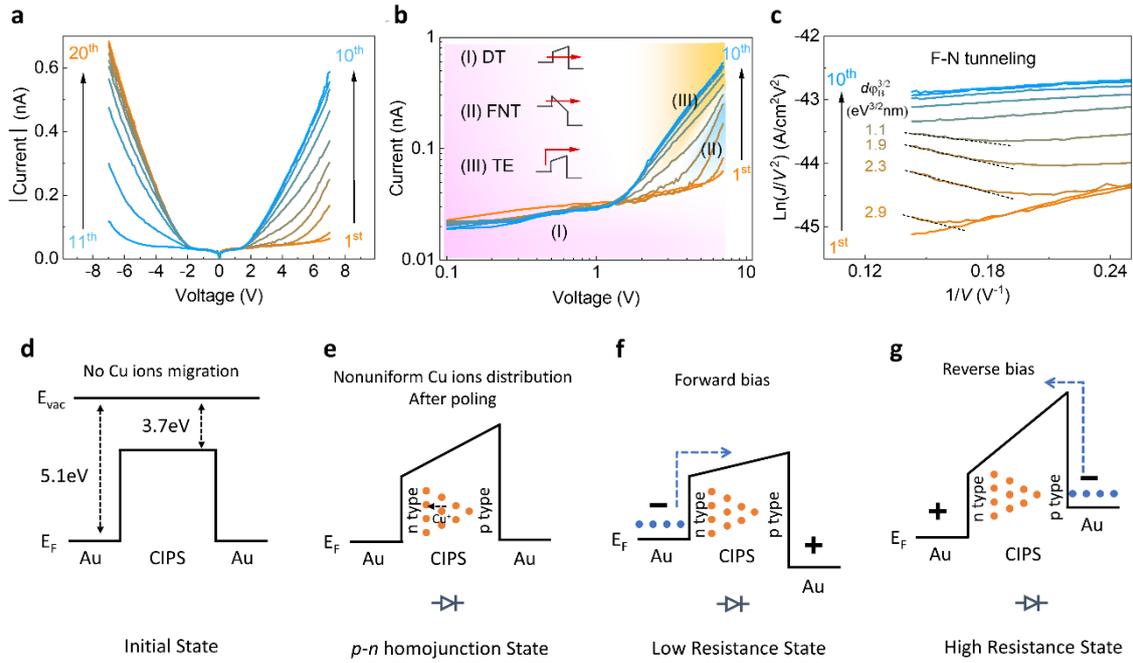

**Figure 4.** Charge transport mechanisms through lateral Au/CIPS/Au homojunctions. (a) *I-V* curves measured with ten consecutive positive/negative voltage sweeps. (b) *I−V* curve on a double logarithmic scale for the positive voltage sweeps. (c) Ln($J/V^2$)-1/$V$ Fowler-Nordheim tunneling plot for the positive voltage sweeps. The order of measurement is labeled as "n$^{th}$". Schematic band alignment of the Au/CIPS/Au device in (d) initial state, and *p-n* homojunction state at (e) zero bias, (f) forward and (g) reverse bias.
17

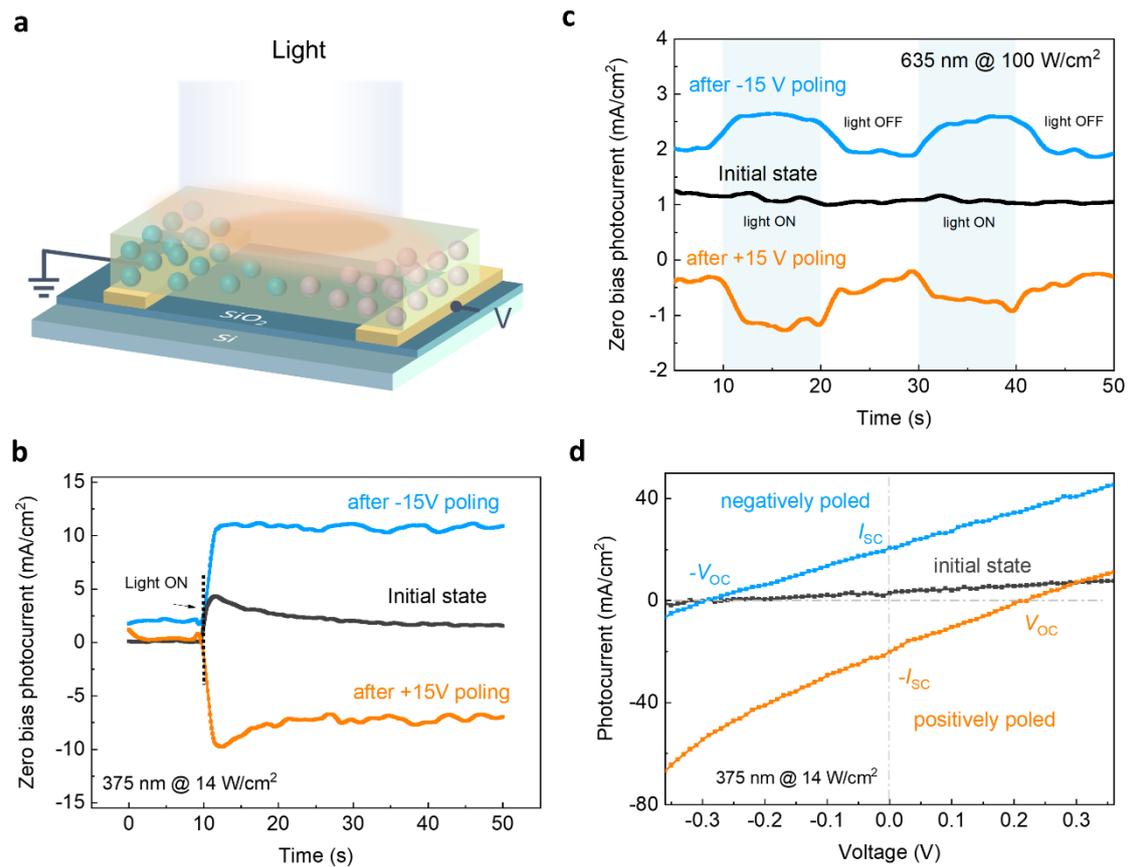

**Figure 5.** Switchable photovoltaic effect in 2D Au/CIPS/Au homojunction. (a) Schematic diagarm of the measurement. (b) Time dependence of the zero bias photocurrents for the device with initial state and after positive/negative poling under an incident light of 375 nm at 14 W/cm$^2$. (c) Temporal dependence of the photocurrent at a zero bias under 635 nm laser illumination at 100 W/cm$^2$. (d) *J–V* characteristics measured with incident light of 375 nm at 14 W/cm$^2$ before and after positive/negative poling.

# Supplementary materials for

**Highly tunable lateral homojunction formed in 2D layered CuInP$_2$S$_6$ *via* in-plane ionic migration**

*Huanfeng Zhu[1,2]\*, Jialin Li[1], Qiang Chen[1], Wei Tang[1], Xinyi Fan[1], Fan Li[1], and Linjun Li[1,2]\**

1. State Key Laboratory of Modern Optical Instrumentation, College of Optical Science and Engineering, Zhejiang University, Hangzhou 310027, China.

2. Intelligent Optics & Photonics Research Center, Jiaxing Research Institute Zhejiang University, Jiaxing 314000, China

*Corresponding Author. E-mail: hfzhu@zju.edu.cn (H.Z.); lilinjun@zju.edu.cn (L.L.)

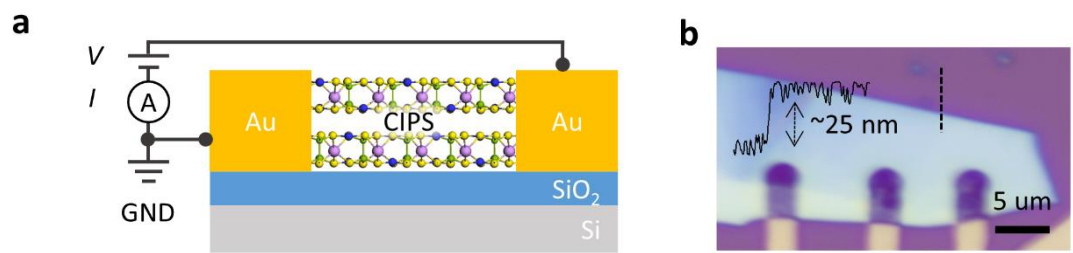

**Figure S1.** (a) The schematic experimental setup for the electrical measurements. (b) The optical images of Au/CIPS/Au device. The thickness of CIPS is 25 nm.

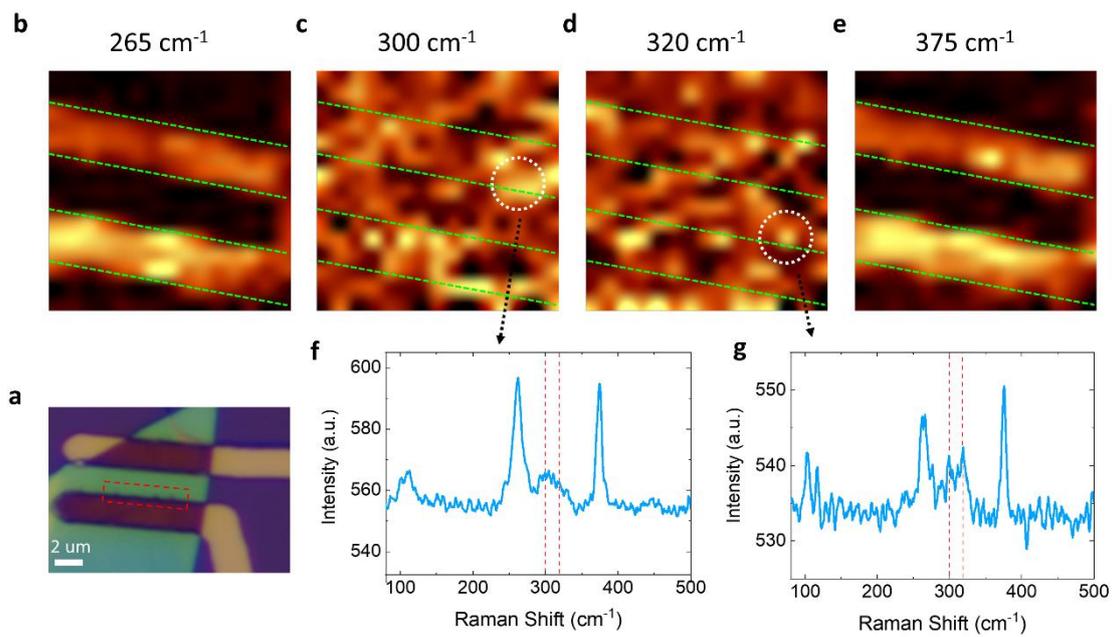

**Figure S2.** (a) Optical microscope image of device after removing the poling voltage. Some dispersed nanoparticles were clearly observed. Raman signal intensity distribution of peak at (b) 265 cm$^{-1}$, (c) 300 cm$^{-1}$, (d) 320 cm$^{-1}$, (e) 375 cm$^{-1}$, respectively. (f), (g) Corresponding Raman spectroscopy in (c) and (e), respectively.

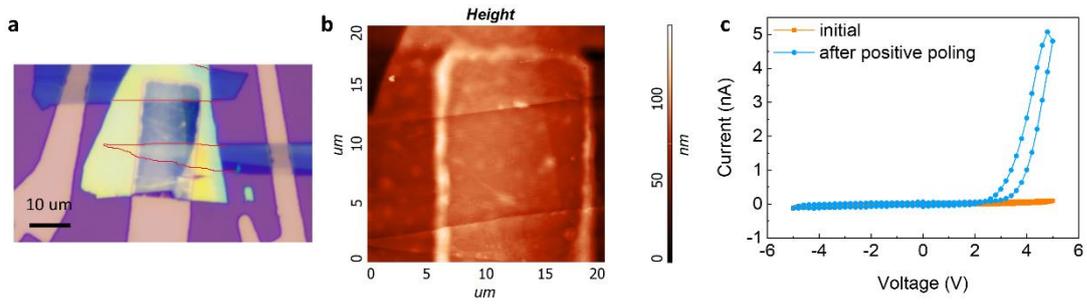

**Figure S3**. The (a) optical and (b) AFM image of the lateral FLG/CIPS/FLG device for the KPFM measurement. (c) I-V curve of the device in the initial and after poling state. Compare with the initial state, after poling processes, the I-V characteristics shows a typical diode-like RS behavior.

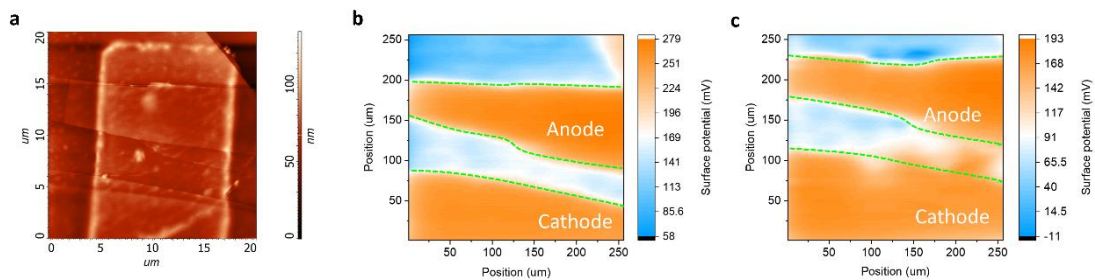

**Figure S4.** (a) The topographic image of the new CIPS device. (b) and (c) The corresponding surface potential mapping of initial and after poling, respectively. The red dotted lines show the approximate positions of the CIPS.

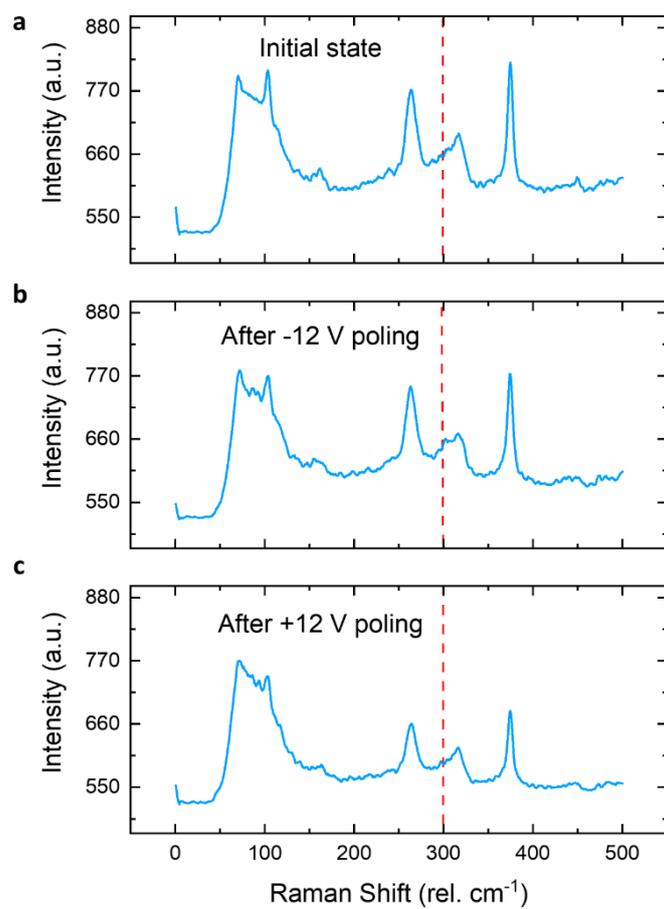

**Figure S5.** Poling voltage-dependent Raman spectroscopy (a) initial state, (b) after -12 V poling and (c) after +12 V poling.

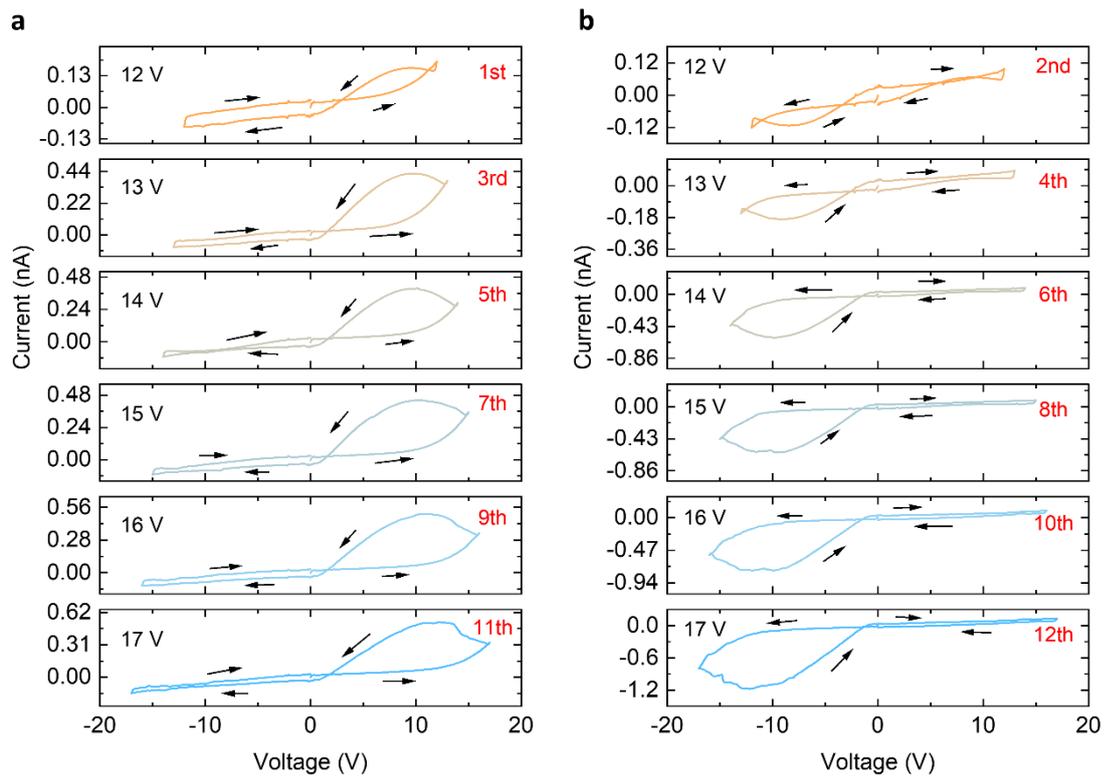

**Fig. S6.** *I–V* curves measured with varying sweep range, Vmax is from 12.0 to 17.0 V. The arrows indicate the voltage sweeping directions. The order of measurement is labeled as "n[th]".

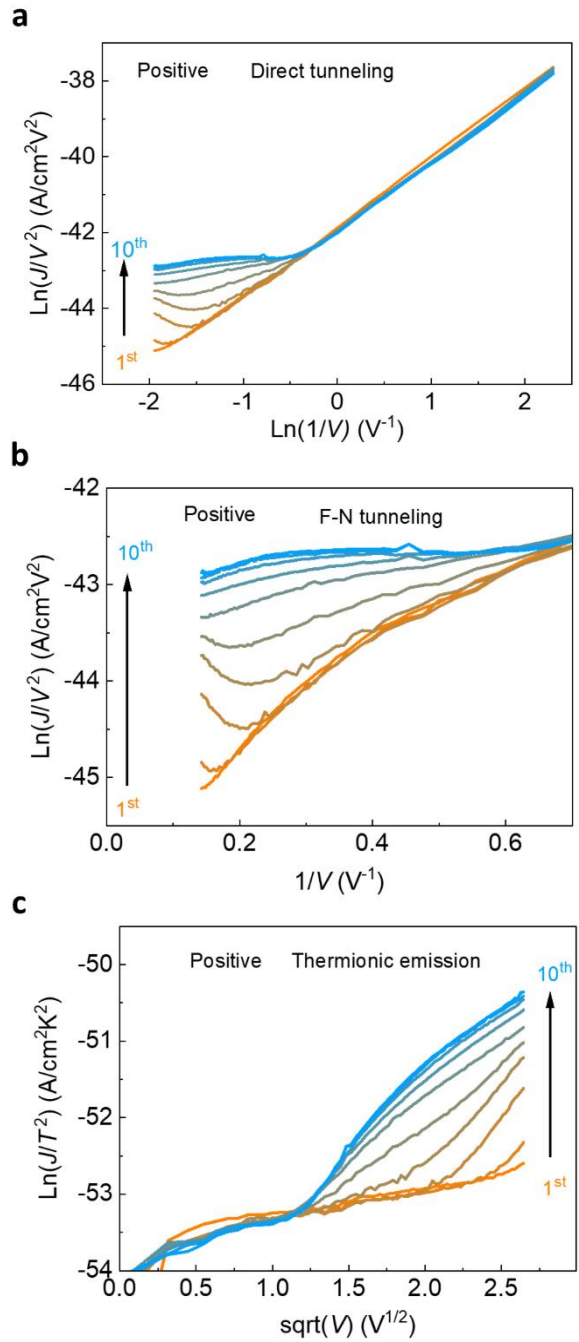

**Figure S7.** *I-V* curves measured with ten consecutive for positive voltage sweeps. (a) Ln($J/T^2$)-Ln($V^{-1}$) direct tunneling plot. (b) Ln($J/V^2$)-$V^{-1}$ Fowler-Nordheim tunneling plot. (c) Ln($J/T^2$)-$V^{1/2}$ thermionic emission plot. The order of measurement is labeled as "n$^{th}$".

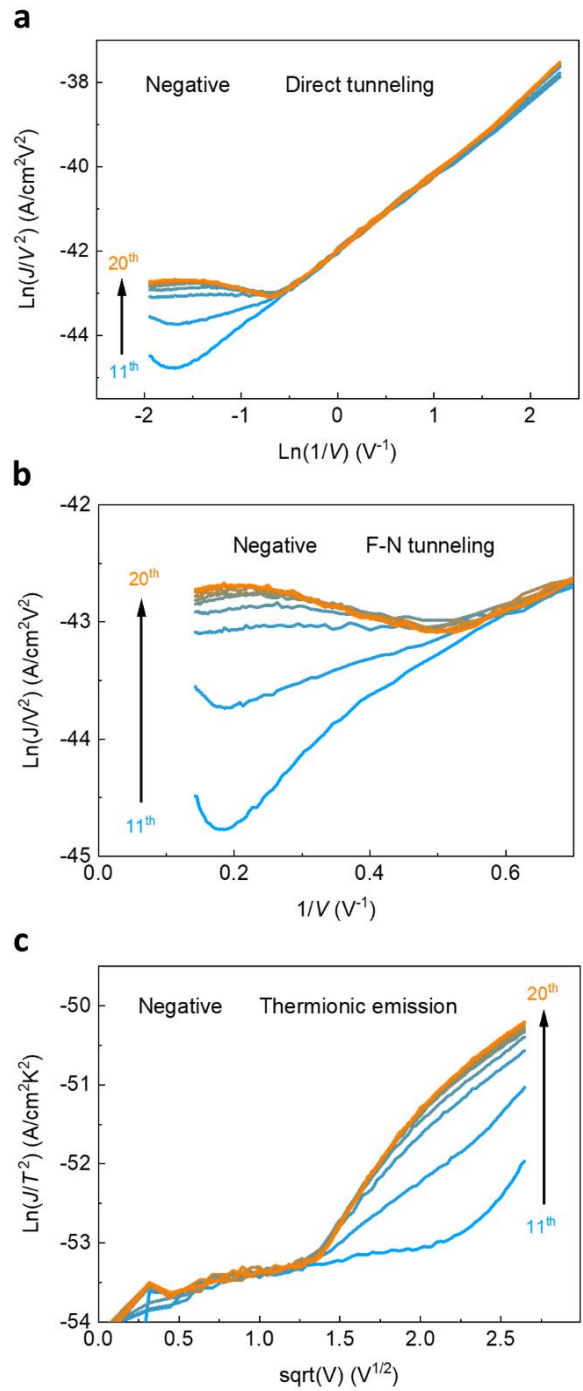

**Figure S8.** *I-V* curves measured with ten consecutive for positive voltage sweeps. (a) Ln($J/T^2$)-Ln($V^{-1}$) direct tunneling plot. (b) Ln($J/V^2$)-$V^{-1}$ Fowler-Nordheim tunneling plot. (c) Ln($J/T^2$)-$V^{1/2}$ thermionic emission plot. The order of measurement is labeled as "n$^{th}$".

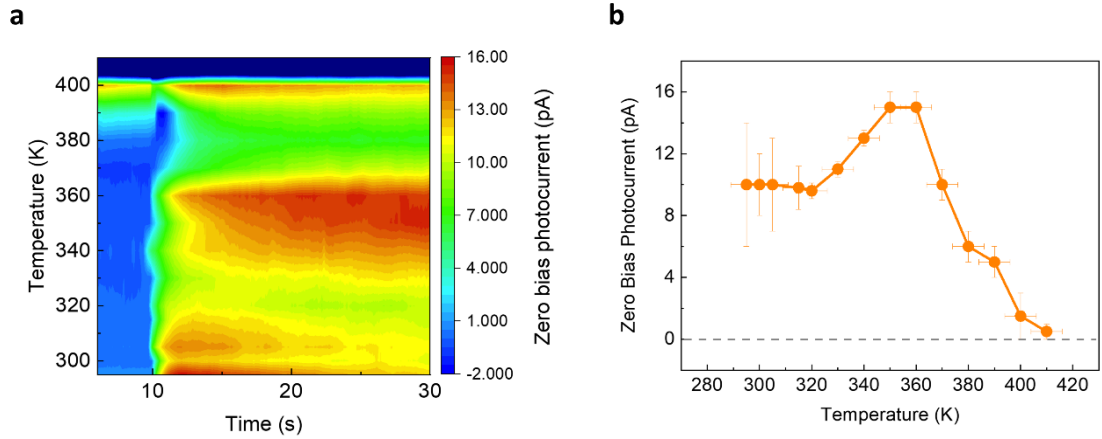

**Figure S9.** (a) Contour plot of zero bias photocurrent vs. temperature and time. (b) The zero bias photocurrent as a function of the temperature extracted from (a). The zero bias photocurrent vanishes when the temperature increases up to 410 K, much higher than the phase transition temperature (~315 K) of CIPS.

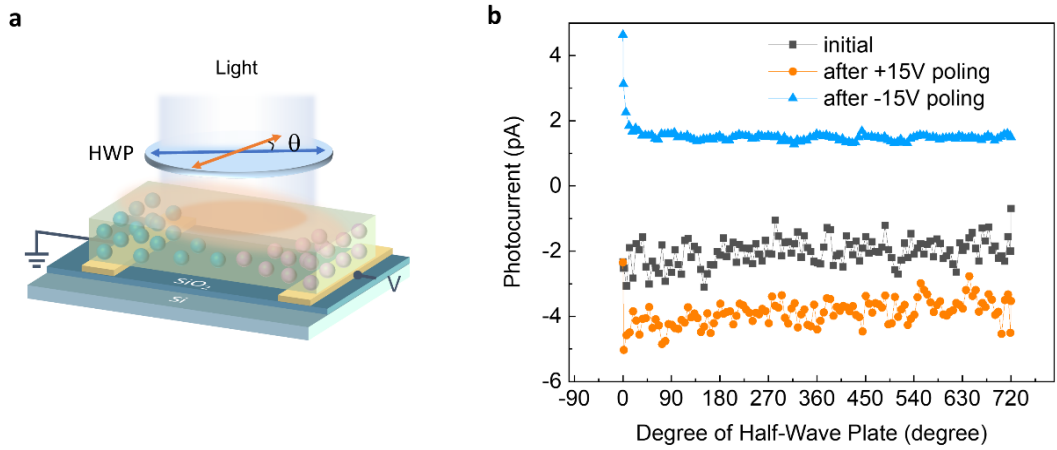

**Figure S10.** (a) The experimental sketch. (b) Angle-dependent zero bias photocurrent. The zero-bias photocurrent is independent of the light polarization, which is inconsistent with the photovoltaic effect scenario.

29